\documentclass
[pra,aps,twocolumn]{revtex4}
\usepackage{amssymb}
\usepackage{latexsym}
\renewcommand{\theequation}{\arabic{section}.\arabic{equation}}
\newcommand{\newsec}{\setcounter{equation}{0}\section}
\newcommand{\Z}{{\mathbb Z}}
\newcommand{\N}{{\mathbb N}}

\newcommand{\C}{{\mathbb C}}
\def\be{\begin{equation}}
\def\ee{\end{equation}}
\def\bea{\begin{eqnarray}}
\def\eea{\end{eqnarray}}
\def\Tr{{\rm \,Tr\,}}

\def\Tr{{\rm \,Tr\,}}

\def\d{{\,\rm d}}

\def\k{{\bf k}}

\def\q{{\bf q}}
\def\n{{\bf n}}
\def\r{{\bf r}}
\def\p{{\bf p}}

\def\sgn{\,{\rm sgn\,}}
\def\npr{{\|\n\|}}
\def\veps{\varepsilon}
\def\h2m{\frac{\hbar^2}{2m}}
\def\p0{{P_{\beta H^0_N}}}

\newtheorem{prop}{Proposition}

\begin{document}

\title{\large\bf Bose-Einstein condensation and symmetry breaking}
\author{Andr\'as S\"ut\H o\\
Research Institute for Solid State Physics and Optics\\ Hungarian Academy of
Sciences \\ P. O. Box 49, H-1525 Budapest\\ Hungary\\
E-mail: suto@szfki.hu}
\thispagestyle{empty}
\begin{abstract}
\noindent
Adding a gauge symmetry breaking field $-\nu\sqrt{V}(a_0+a_0^*)$ to the
Hamiltonian of some simplified models of an interacting Bose gas we compute
the condensate density and the symmetry breaking order parameter in the limit of
infinite volume and prove Bogoliubov's asymptotic hypothesis
$
\lim_{V\to\infty}\langle a_0\rangle/\sqrt{V}={\rm sgn\,}\nu\,
\lim_{V\to\infty}\sqrt{\langle a_0^*a_0\rangle/V}
$
where the averages are taken in the ground state or in thermal equilibrium states.
Letting $\nu$ tend to zero in this equation we obtain
that Bose-Einstein condensation occurs if and only if the gauge symmetry is
spontaneously broken. The simplification consists in dropping the off-diagonal
terms in the momentum representation of the pair interaction.
The models include the mean field and the imperfect (Huang-Yang-Luttinger) Bose gas.
An implication of the result is that the compressibility sum rule cannot hold true in the
ground state of the one-dimensional mean-field Bose gas.
Our method is based on a resolution of the Hamiltonian into a family of single-mode
($\k=0$) Hamiltonians and on the analysis of the associated microcanonical ensembles.

\vspace{2mm}

\vspace{2mm}
\noindent

\end{abstract}
\maketitle
\newsec{Introduction and results}

The relation between Bose-Einstein condensation (BEC) and a
spontaneous breakdown of the gauge symmetry of the system
(shortly, symmetry breaking) is not entirely understood. Symmetry
breaking is absent in textbook discussions of BEC of the ideal
Bose gas and does not appear in the general definition of BEC by
Penrose and Onsager \cite {PO} either. It is, on the other hand, a
central notion of Bogoliubov's theory of superfluidity and
superconductivity, dealt with extensively under the name of
quasi-averaging \cite{Bog}.
Bose-Einstein condensation is the
accumulation of a macroscopic number of particles in a
single-particle state. In homogenous (translation invariant)
situation this state is $\varphi_0(\r)\equiv e^{i\phi}/\sqrt{V}$
where $V$ is the volume of the cube on which the system is defined
with periodic boundary conditions and $\phi$ is a real constant.
The macroscopic occupation of $\varphi_0$ means that the thermal
(or ground state) average of its occupation number operator $N_0$
increases proportionally to $V$ as $V$ tends to infinity, which
can be shown to be equivalent
to an off-diagonal long-range order (ODLRO). 
Hamiltonians conserving particle number possess a continuous gauge
symmetry related to the phase $\phi$. For pair interactions a
general Hamiltonian reads
\bea\label{1}
H=\sum\veps(\k) N_k-\mu N
+\frac{1}{2V}\sum_{\q,\k,\k'}v(\q)
a^*_{\k+\q}a^*_{\k'-\q}a_{\k'}a_{\k}\nonumber\\
\eea
where
$\veps(\k)=\hbar^2\k^2/2m$, $\mu$ and $v(\q)=v(-\q)$ are real numbers,
$N_\k=a^*_{\k}a_{\k}$, $N=\sum N_\k$, $a^*_{\k}$ creates a boson
in the single-particle state
$\varphi_\k(\r)=(1/\sqrt{V})\exp\{i\k\r\}$ and $\q,\k,\k'$ are
allowed wave vectors, i.e. compatible with the periodic boundary
condition. Averaging any $a_\k$ with eigenstates or the density
matrix generated by $H$ yields zero. (We disregard a possible
accidental degeneracy of eigenstates belonging to different
particle numbers.) For any real $\phi$ changing all $a_\k$
simultaneously into $e^{-i\phi}a_\k$ (and $a^*_\k$ into
$e^{i\phi}a^*_\k$) leaves the commutation relations and the
Hamiltonian invariant. This gauge symmetry can be broken by adding
a term which breaks particle number conservation, e.g. for real
$\nu$ and $\phi$ the operators
$
H-\nu\sqrt{V}\left(e^{-i\phi}a_0+e^{i\phi}a^*_0\right)
$
are no more gauge invariant. Still they are unitary equivalent
for different $\phi$, therefore $\phi=0$ can be taken without
restricting generality. Now the thermal or ground state
average
$\langle a_0\rangle$ is non-vanishing, and we speak about
a spontaneous breakdown of the gauge symmetry if the
`quasi-average'
$$
\lim_{\nu\to 0}\lim_{V\to\infty}\frac{1}{\sqrt{V}}\langle
a_0\rangle\neq 0\ .
$$
When this occurs, we are dealing with a continuum of infinite
volume pure states, each characterized by a phase $\phi$, and
the state we obtain by fixing $\nu=0$ in finite volumes and
taking the limit $V\to\infty$ is the uniform mixture of
pure states with $0\leq\phi<2\pi$.

By now the physical reality of symmetry breaking
-- at least in a non-translation-invariant situation --
has become an experimental fact, manifesting itself in the interference of
two condensates of different phases \cite{Ket}.
It is all the more embarrassing that there seems to exist no
theoretical proof of its occurrence. There is
Hohenberg's theorem on the {\em absence} of symmetry breaking in
one and two dimensional Bose gases at positive temperatures
\cite{Ho,BM,rem}. In some other cases one can prove BEC by showing
either the existence of ODLRO or the macroscopic occupation of a
one-particle state. An example for the first is the hard-core Bose
lattice gas at half-filling \cite{DLS}, for the second the trapped
Bose gas \cite{Su}. However, the Schwarz inequality
\be\label{4} |\langle
a_0\rangle|\leq \sqrt{\langle N_0\rangle}
\ee
indicates that the absence of symmetry breaking does not automatically
imply the absence of
BEC, and {\em vice versa}, BEC does not obviously imply symmetry breaking.
A closely related question is the validity of
Bogoliubov's approximation \cite{BogA} which consists in replacing $a_0$
and $a^*_0$ by complex numbers $c$ resp. $c^*$.
A partial answer was given by Ginibre a long time ago \cite{Gin}.
Ginibre considered this replacement in the Hamiltonian or in the
grand-canonical density matrix, with or without the symmetry breaking field, and
proved that in all cases by choosing $c$ so that it maximizes the approximate
pressure in finite volumes the result
converges to the exact pressure in the limit of infinite volume. This
suggests that symmetry breaking indeed takes place when there
is BEC, and the inequality (\ref{4}) saturates asymptotically.

In this paper we will prove that for any $\nu\neq 0$,
\be\label{Bogogen}
\lim_{V\to\infty}\langle a_0\rangle/\sqrt{V}={\rm sgn\,}\nu\,
\lim_{V\to\infty}\sqrt{\langle N_0\rangle/V}
\ee
in the ground state
and thermal equilibrium states
of the mean-field Bose gas and of other models whose
interaction is diagonal in momentum representation.
The corresponding Hamiltonians are obtained from
(\ref{1}) by dropping the off-diagonal terms:
\bea\label{FD}
H^{\rm FD}=\sum\veps(\k) N_k-\mu N+\frac{v_0}{2V}(N^2-N)\nonumber\\
+\frac{1}{2V}\sum_{\k\neq\k'}v(\k'-\k)N_\k N_{\k'}\ .
\eea
Here
$v_0=v(0)>0$, and for the sake of simplicity we consider only $v(\k)\geq 0$.
In the mean-field model, $H^{\rm MF}$,
$v(\k)=0$ for $\k\neq 0$. The imperfect Bose gas
is defined by keeping the diagonal part of the $\delta$ interaction, so that
$v(\k)=v_0$ for all $\k$ \cite{HYL}.
Note that only the interaction of the mean-field Bose gas is diagonal in coordinate representation. The
other models may present nonintuitive features like anomalies in the spectrum of elementary excitations,
see below, and
the Thouless effect, a jump-discontinuity in the
condensate density as the chemical potential increases \cite{Th,DLP}.
The thermal equilibrium properties of the full diagonal model (\ref{FD}) were rigorously studied
by Dorlas et al. \cite{DLP}.

The ground state of each Hamiltonian of the family (\ref{FD}) in the Fock space
is $|N=N_0=m_V\rangle$ where $m_V$
is the nonnegative integer closest to $\mu V/v_0+\frac{1}{2}$.
In the limit of infinite volume
the ground state energy density
is $-(v_0/2)\lim(m_V/V)^2$.
We will study the Hamiltonian
\be
H^{\rm FD}_\nu=H^{\rm FD}-\nu\sqrt{V}(a_0+a^*_0)
\ee
and prove that for $\nu\neq 0$ the ground state energy density
\bea\label{8}
e_0\equiv\lim_{V\to\infty}E_0/V
&=&\min_{s\geq 0}\{v_0s^2/2-\mu s-2|\nu|\sqrt{s}\,\}\nonumber\\
&=&v_0s_0^2/2-\mu s_0-2|\nu|\sqrt{s_0}\ ,
\eea
the order parameter
\be\label{7}
\alpha_0\equiv
\lim_{V\to\infty}\frac{1}{\sqrt{V}}\langle a_0\rangle={\rm sgn\,}\nu\,\sqrt{s_0}
\ee
and the condensate density
\be\label{ro0}
\rho_0\equiv
\lim_{V\to\infty}\frac{1}{V}\langle N_0\rangle=s_0=\alpha_0^2\ .
\ee
In (\ref{7}) and (\ref{ro0}) the averages are taken in the ground state of
$H^{\rm FD}_\nu$. The minimizer $s_0$ coincides with the full particle density.
It is positive for any value of $\mu$, but
goes to zero with a vanishing $\nu$ if $\mu\leq 0$.
The extension of Eq.~(\ref{Bogogen}) to positive temperatures
will be obtained by proving the analog of
Eqs.~(\ref{8})-(\ref{ro0}) for the ground state of certain microcanonical
ensembles and by referring to the equivalence of micro- and grand-canonical
ensembles. Equation (\ref{Bogogen}) is not a statement about the existence
of BEC. It holds even if the chemical potential is below its critical value when
both sides vanish as $\nu$ tends to zero. It however guaranties that BEC and a
spontaneous breakdown of the gauge symmetry occur simultaneously.

The result for the ground state is independent of the dimension, although
$H^{\rm FD}$ depends on it. In particular, there is symmetry breaking along with the
Bose condensation in the ground state of the one-dimensional mean-field
Bose gas. Because of the finite compressibility of this system, our finding is
in conflict
with an earlier argument according to which there should be no symmetry breaking
in the ground state of one-dimensional interacting Bose gases \cite{PS}.
To track down the source of this contradiction and also by a general interest
we consider here some implications
of the results listed above on the properties of the excitation spectrum.
The non-relativistic version of the Goldstone theorem states that a spontaneously broken continuous
symmetry implies the existence of gapless excitations, i.e. eigenstates whose energy tends to zero
with a vanishing momentum. Wagner's
analysis \cite{Wag} based on the Bogoliubov inequality and on Bogoliubov's $1/q^2$-theorem concluded that
the nature of the long-wavelength excitations was uncorrelated with symmetry breaking and was
governed by the range of the interaction. Our conclusion for this particular family of interactions is
similar. As it is shown below, there are gapless excitations in the mean-field model whose interaction
is of infinite range but there is a gap to excitations with a non-vanishing momentum
in the spectrum of the other models.
This seems to lead to another contradiction.
In the ground state version of the $1/q^2$-theorem,
\be\label{9}
\langle N_\q\rangle\geq\frac{1}{4}\
\frac{(E_\q-E_0)\, |\langle a_0\rangle|^2}{\veps(\q)\,
\langle N\rangle+|\nu\langle a_0\rangle|\sqrt{V}}-\frac{1}{2}\ ,
\ee
$E_0$ is the ground state energy and $E_\q$ is the
energy of the lowest lying eigenstate of momentum $\hbar\q$ of the symmetry breaking
Hamiltonian \cite{remark}.
In the actual case (\ref{9}) can be made stronger.
The Hamiltonian $H^{\rm FD}_\nu$ commutes separately with each $N_\k$, $\k\neq 0$.
We shall prove that for $\nu\neq 0$
the ground state is still in the invariant subspace $\{N_\k=0|\k\neq 0\}$.
Then the ground state average $\langle N_\q\rangle=0$ and from the derivation
of Eq.~(\ref{9}) one sees that
$E_\q$ can be taken to be the smallest eigenvalue among the eigenstates
with $N_\q=1$ and $N_\k=0$ if $\k\neq 0$ or $\q$, i.e. a single particle
carrying the total momentum, which can be much greater than the
minimum eigenvalue among eigenstates of momentum $\hbar\q$. In general,
summing (\ref{9}) over allowed $\q$ such that $0<|\q|<q_0$, dividing by $V$ and
letting $V$ tend to infinity while keeping $q_0$ fixed, we arrive at
\bea\label{sym}
\rho-\rho_0&\geq& \frac{1}{2(2\pi)^d}\nonumber\\
&\times&\int_{|\q|<q_0}\left(\frac{1}{2}\frac{|\alpha_0|^2\lim_{V\to\infty}(E_\q-E_0)}
{\veps(\q)\rho+|\nu \alpha_0|}-1\right)\d\q\nonumber\\
\eea
where $\rho$ is the total particle density and $d$ is the dimension~\cite{rem2}.
In the actual case $\rho=\rho_0=|\alpha_0|^2$.
In Section \ref{En} we will derive the gap to one-particle excitations,
\bea\label{10}
E_\q-E_0=\veps(\q)+\mu\frac{v(\q)}{v_0}+|\nu|\frac{v_0+v(\q)}{\sqrt{\mu v_0}}
\nonumber\\
+O(\nu^2)+O(1/V),
\eea
valid for $\mu>0$.
The finite-size correction $\sim 1/V$ is uniform in $\nu$ so that the limits $V\to\infty$ and $\nu\to 0$
are interchangeable.
We can use this result in (\ref{sym}) and see what we obtain when $\nu$ goes to zero.
For the mean-field model
$\lim_{\nu\to 0}\lim_{V\to\infty}(E_\q-E_0)=\veps(\q)$, therefore
(\ref{sym}) holds true.
For the other models the gap implies a diverging integral in one and two
dimensions and, thus, a violation of the inequality (\ref{sym}).
The cause of the contradiction is our misuse of
(\ref{9}), which cannot be applied to $H^{\rm FD}_\nu$ if $v(\k)\neq 0$ for some $\k\neq 0$:
These interactions are non-diagonal (nonlocal)
in coordinate representation and therefore do not commute with $\rho_\k$, the Fourier transform of the
density operator, which is diagonal. This
invalidates the $f$-sum rule \cite{PN}, used in the derivation of Eq.~(\ref{9}).
Turning to the controversy in connection with the mean-field model, we recall
an improved version of (\ref{9}) due to Pitaevskii and
Stringari~\cite{PS},
\be\label{PitS}
\langle N_\q\rangle\geq\frac{1}{4}\
\frac{|\langle a_0\rangle|^2}{\langle\rho_{-\q}\rho_\q\rangle}-\frac{1}{2}
=
\frac{1}{4}\
\frac{|\langle a_0\rangle|^2}{\langle N\rangle S(\q)}-\frac{1}{2}
\ee
where $S(\q)$ is the static structure function. This formula is
valid for all the models discussed in this paper and leads to
\be
\int_{|\q|<q_0}\left(\frac{1}{2S(\q)}-1\right)\d\q\leq 0\ .
\ee
We can conclude that $q^d/S(\q)\to 0$ as $q\to 0$ in all models considered here.
In the general case, when the $f$-sum rule holds true, (\ref{9}) can be
obtained also from (\ref{PitS}). When both the $f$- and the compressibility sum rules~\cite{PN}
hold,
\be\label{compress}
\lim_{q\to 0}q/S(\q)\geq \sqrt{4m/\hbar^2\kappa\rho},
\ee
where $\kappa$ is
the ground state compressibility. In the mean-field Bose gas
$\kappa=1/v_0\rho^2<\infty$. Thus, (\ref{compress}) cannot hold true in one
dimension and because
the $f$-sum rule is valid, the compressibility sum
rule must fail in the ground state of the one-dimensional mean-field
Bose gas.

Note that the positive temperature variant of inequality (\ref{9}),
with $2k_BT$ replacing $E_\q-E_0$, cf.~\cite{Bog,remark},
also relies on the commutability of $\rho_\k$ with the interaction. Thus, it
cannot be applied to prove the absence of symmetry breaking for $H^{\rm FD}$ in one and
two dimensions at $T>0$
if $v(\k)$ is nonvanishing for nonzero $\k$. Rightly so, because
in these systems, e.g. in the imperfect Bose gas \cite{HYL},
there can be BEC at $T>0$ even in one dimension, see also \cite{Th,DLP}.

All our results are obtained by representing the Hamiltonian as a Jacobi
matrix in microcanonical ensembles or,
equivalently, by studying some tight-binding Schr\"odinger
equations on the spectrum of $N_0$.
In Section \ref{Gs} we show that $N_\k=0$ for all $\k\neq 0$
in finite volume ground states.
Formulas (\ref{8})-(\ref{ro0}) will be derived in Section \ref{Gsed}.
The analog of (\ref{8}) for the ground state of microcanonical ensembles is presented
there, that of (\ref{7}) and (\ref{ro0}) in Section \ref{En} which also contains
the derivation of Eq.~(\ref{10}). In Section \ref{pos}
we discuss the extension of the results to positive temperatures. The paper ends
with a Summary.

\newsec{Ground state in finite volumes}\label{Gs}

The Hamiltonians
commute with each $N_\k$, $\k\neq 0$. One can therefore decompose them
into operators acting on invariant subspaces with fixed $N_\k$. Let $\n=\{n_\k\}_{\k\neq 0}$ denote
terminating sequences of nonnegative integers, i.e. $\npr=\sum_{\k\neq 0}n_\k<\infty$. Then
\be\label{resoln}
H^{\rm FD}_\nu=\oplus_{\n}H(\n)
\ee
where $H(\n)$ is the restriction of $H^{\rm FD}_\nu$ onto the subspace $\{N_\k=n_\k|\k\neq 0\}$.
Because $v_0>0$, $H^{\rm FD}_\nu$ has a discrete point spectrum with $+\infty$ as the only
accumulation point, $\Tr \exp\{-\beta H^{\rm FD}_\nu\}<\infty$ for $\beta>0$,
and the same holds for each $H(\n)$.
In this section we prove the following assertion.

\vspace{2mm}
\noindent
{\em
In any finite volume the ground state of $H^{\rm FD}_\nu$ is in the subspace
$\{N_\k=0|\k\neq 0\}$, i.e. it is the ground state of $H(0)$.
}

\vspace{2mm}
\noindent
This property is specific to the diagonal models and $v(\k)\geq 0$. A general interaction does not commute
with $N_\k$. This knowledge about the ground state can but need not be used for the
derivation of the formulas (\ref{8}), (\ref{7}) and (\ref{ro0}). It plays, however, a role in the
interpretation and computation of $E_\q-E_0$ in Eq.~(\ref{10}).

One has to prove that for each $\n\neq 0$ the ground state energies of $H(\n)$ and $H(0)$
satisfy the inequality
$E_0[H(\n)]> E_0[H(0)]$. This is immediately seen if
$\npr\geq 1+2\mu V/v_0$. The operator $H(\n)-H(0)$ is diagonal and in this case
it has positive elements,  
which implies the result. In the case $\npr<1+2\mu V/v_0$ (which can occur only
if $\mu>0$), however, it may have
both positive and negative elements and we need a more elaborate argument.
We shall interpose
between $H(\n)$ and $H(0)$ three auxiliary operators, $K(\npr)$, $K(\npr,0)$ and $K(\npr,1)$ and show the
following order of ground state energies:
\bea\label{2.2}
E_0[H(\n)]>E_0[K(\npr)]
=E_0[K(\npr,0)]\nonumber\\
> E_0[K(\npr,1)]\geq E_0[H(0)]\ .
\eea
Define
\be\label{2.3}
K(p)=\frac{v_0}{2V}(N_0+p)^2-\left(\mu+\frac{v_0}{2V}\right)(N_0+p)
-\nu\sqrt{V}(a_0+a_0^*)
\ee
where $p$ is a real number.
In particular, $K(0)=H(0)$. The operator inequality
\be\label{H>K}
H(\n)> K(\npr)
\ee
holds true for $\n\neq 0$, because $H(\n)-K(\npr)$ is a diagonal operator with entries
$\geq\sum\veps(\k) n_\k>0$.
For the ground state energies (\ref{H>K}) implies
\be\label{EH>EK}
E_0[H(\n)]> E_0[K(\npr)]\ .
\ee
Let $p$ be a nonnegative integer.
Using the basis $\{|N_0=m\rangle\}_{m=0}^\infty$ of $N_0$-eigenstates, $K(p)$ can be represented
by a semi-infinite tridiagonal matrix (Jacobi matrix), denoted also by $K(p)$.
The diagonal matrix elements satisfy
$K(p)_{mm}=K(0)_{m+p,m+p}$ for $m\geq0$. The off-diagonal elements are the same for all $p$,
$K(p)_{m-1,m}=K(p)_{m,m-1}=-\nu\sqrt{mV}$ if $m\geq 1$, and 0 otherwise.
The matrices $K(p)$ act in the Hilbert space
$\ell_2(\N)=\{(\psi(m))_0^\infty|\sum_{m}|\psi(m)|^2<\infty\}$. We extend them without changing the
notation onto
$\ell_2(\Z)=\{(\psi(m))_{-\infty}^\infty|\sum_{m}|\psi(m)|^2<\infty\}$
by setting $K(p)_{mn}=0$ if either $m$ or $n$ is negative. Then $\ell_2(\N)$ and $\ell_2(\Z\setminus\N)$
are invariant subspaces of each $K(p)$ which acts on the first space as the original operator and on the
second as the null operator. If $E_0[K(p)]<0$, the ground state energy of the extended operator is
the same as that of the original one. Let $U_p$ be the left shift by $p$, i.e. for $\psi\in\ell_2(\Z)$,
$
(U_p\psi)(m)=\psi(m+p)
$.
Then
\bea
(U_p^{-1}K(p)U_p)_{mm}=K(p)_{m-p,m-p}\nonumber\\
=\left\{\begin{array}{cl}
                          0&,\quad m\leq p-1\\
                          K(0)_{mm}&,\quad m\geq p
                              \end{array}\right.
\eea
and
\bea
(U_p^{-1}K(p)U_p)_{m-1,m}=K(p)_{m-1-p,m-p}\nonumber\\
=\left\{\begin{array}{cl}0 &,\quad m\leq p\\
                                               -\nu\sqrt{(m-p)V}&,\quad m\geq p+1
                              \end{array}\right.\ .
\eea
Define $K(p,0)=U_p^{-1}K(p)U_p$. The operator $U_p$ is unitary, thus $E_0[K(p,0)]=E_0[K(p)]$.

For the definition of $K(p,1)$ we split the matrices $K(p,0)$ and $K(0)$ into diagonal $D$ and off-diagonal $Q$ parts,
$K(p,0)=D(p)+Q(p)$ and $K(0)=D(0)+Q(0)$, and set $K(p,1)=D(p)+Q(0)$. Furthermore, we introduce an
interpolating matrix function
\be
K(p,t)=tK(p,1)+(1-t)K(p,0)=D(p)+t[Q(0)-Q(p)]\ .
\ee
By the Hellmann-Feynman theorem
\bea\label{difft}
\frac{\partial}{\partial t}E_0[K(p,t)]
=(\psi_t,[Q(0)-Q(p)]\psi_t)\nonumber\\
=\sum_{m,n\geq 0} \psi_t(m)[Q(0)-Q(p)]_{mn}\psi_t(n)
\eea
where $\psi_t$ is the normalized ground state of $K(p,t)$. Choose $\nu$ to be positive (changing the
sign of $\nu$ amounts to a unitary transformation with a diagonal matrix of elements $(-1)^m$).
Then
$$[Q(0)-Q(p)]_{m-1,m}=[Q(0)-Q(p)]_{m,m-1}<0$$
for $m\geq 1$ implying (through the variational
principle or the Perron-Frobenius theorem)
$\psi_t(m)>0$ for $m\geq 0$. Thus the derivative (\ref{difft}) is negative, which proves
$E_0[K(p,0)]>E_0[K(p,1)]$.

At last, we note that $K(p,1)-K(0)=D(p)-D(0)\geq 0$ if $p=\npr\leq 1+2\mu V/v_0$ because
\be
[D(p)-D(0)]_{mm}=\left\{\begin{array}{cl}
                                         -K(0)_{mm}\geq 0 &,\quad 0\leq m\leq p-1\\
                      0 &,\quad {\rm otherwise}\ .
                        \end{array}\right.
\ee
Therefore $E_0[K(\npr,1)]\geq E_0[K(0)]$ indeed. Here $E_0[K(0)]$ is the ground state energy of the
extended operator. If $E_0[H(0)]$ was positive, we would obtain $E_0[K(0)]=0$. However,
$E_0[H(0)]<E_0[H^{\rm FD}]<0$ can be seen by using $|N=N_0=m_V\rangle$ as a variational wave function,
so that $E_0[K(0)]=E_0[H(0)]$.

\newsec{Ground state in infinite volume}\label{Gsed}

Our aim in this section is to derive Eqs. (\ref{8})-(\ref{ro0}).
Incidentally, we shall
obtain the ground state energy density for $H(\n)$, cf. (\ref{resoln}).

Each $H(\n)$ is an unbounded operator which is bounded below. A common lower bound is
that of $H^{\rm FD}_\nu$,
\bea\label{27}
H^{\rm FD}_\nu&\geq& H^{\rm FD}-|\nu|\left(V+N_0\right)\nonumber\\
&\geq&
-\left[\frac{v_0}{2}\left(\frac{\mu+|\nu|}{v_0}+\frac{1}{2V}\right)^2 +|\nu|\right]V.
\eea
Here we have made use of
$
\pm\sqrt{V}(a_0+a_0^*)\leq V+N_0
$
which comes from
$
(\sqrt{V}\pm a_0^*)(\sqrt{V}\pm a_0)\geq 0
$.

To obtain a good variational upper bound on $E_0[H(\n)]$, note that $H^{\rm FD}$ of Eq. (\ref{FD})
has a decomposition analogous to (\ref{resoln}),
$H^{\rm FD}=\oplus_\n H_{\rm diag}(\n)$
where
$H_{\rm diag}(\n)$ is the diagonal part of $H(\n)$. Let $|N_0=J(\n)\rangle$ be the ground state of
$H_{\rm diag}(\n)$ ($J(0)=m_V$). Then with
$$\psi=V^{-1/4}\sum_{m=J(\n)}^{J(\n)+\sqrt{V}}|N_0=m\rangle$$
as a variational wave function
we obtain
\be
E_0[H(\n)]\leq E_0[H_{\rm diag}(\n)]-2|\nu|\sqrt{J(\n)V}+O(\sqrt{V})\ .
\ee
This upper bound will turn out to be the precise asymptotic result up to order $|\nu|$.

Let us introduce
\bea\label{xyz}
x_V(\n)=\frac{1}{V}\sum_{\k\neq0}n_\k\qquad
y_V(\n)=\frac{1}{V}\sum_{\k\neq 0}v(\k) n_\k \qquad\nonumber\\
z_V(\n)=\frac{1}{V}\sum\veps(\k) n_\k+\frac{1}{2V^2}
\sum_{0\neq\k\neq\k'\neq 0}v(\k'-\k)n_\k n_{\k'}.\nonumber\\
\eea
All three are real nonnegative numbers, vanishing if $\npr=0$. With them
\bea\label{hV}
H(\n)/V\equiv
h_V(x_V,y_V,z_V)\phantom{aaaaaaaaaaaaaaaaaaaaaaa}\nonumber\\
=\frac{v_0}{2}\left(\frac{N_0}{V}+x_V\right)^2
-\left(\mu+\frac{v_0}{2V}\right)
\left(\frac{N_0}{V}+x_V\right)\nonumber\\
+y_V\frac{N_0}{V}-\frac{\nu}{\sqrt{V}}(a_0+a_0^*)+z_V\ .
\nonumber\\
\eea
As in the previous section, we consider the matrix of each $h_V$ in the basis of the $N_0$-eigenstates and
extend it to $\ell_2(\Z)$ with zero elements for negative indices.
The nonvanishing matrix elements of $h_V$ are
\bea
h_V(x_V,y_V,z_V)_{jj}=\phantom{aaaaaaaaaaaaaaaaaaaaaaaaaaaa}
\nonumber\\
\frac{v_0}{2}\left(\frac{j}{V}+x_V\right)^2-\left(\mu+\frac{v_0}{2V}\right)
\left(\frac{j}{V}+x_V\right)+y_V\frac{j}{V}+z_V\nonumber\\
\eea
and
\be
h_V(x_V,y_V,z_V)_{j,j+1}=h_V(x_V,y_V,z_V)_{j+1,j}=-\nu\sqrt{\frac{j+1}{V}}
\ee
with $j\geq 0$.

Each $h_V$ is still an unbounded operator. 
We can zoom in on different parts of its spectrum
by employing different $V$-dependent {\em right} shifts in $\ell_2(\Z)$
and taking the limit $V\to\infty$.
(Left shifts would only generate the zero matrix.) Suppose that
$x_V\to x$, $y_V\to y$ and $z_V\to z$. The limits $x$, $y$ and $z$ may not be
arbitrary nonnegative numbers, e.g. $y=0$ for the mean-field model and $y=v_0x$
for the imperfect Bose gas. 
Choose a sequence
$S_V$ of positive integers such that $s_V=S_V/V\to s$.
For the matrix elements the shift corresponds to a change of
variables, $j=m+S_V$, yielding
$h_V(x_V,y_V,z_V,s_V)_{mm'}=h_V(x_V,y_V,z_V)_{m+S_V,m'+S_V}$. When $V$ tends to infinity
the limit of $h_V(x_V,y_V,z_V,s_V)_{mm'}$
exists and defines a doubly infinite Jacobi matrix $h(x,y,z,s)$ with nonvanishing elements~\cite{rem3}
\bea\label{diag}
h(x,y,z,s)_{mm}\equiv \chi(x,y,z,s)\phantom{aaaaaaaaaaaaaa}\nonumber\\
=\frac{v_0}{2}(s+x)^2-\mu(s+x)+ys+z
\eea
and
\be
h(x,y,z,s)_{m,m+1}=h(x,y,z,s)_{m+1,m}=-\nu\sqrt{s}\ .
\ee
Formally $h(x,y,z,s)$ is a one-dimensional tight-binding Schr\"odinger operator
with a constant potential (\ref{diag}), governing the motion of a particle of mass
$\sim 1/|\nu|\sqrt{s}$ along the spectrum of $N_0$.
Physically $h(x,y,z,s)$ is associated with the infinite volume limit of a
microcanonical ensemble of bosons
characterized by four intensive variables:
the condensate density $s$,
the density $x$ and energy density $-\mu x+v_0x^2/2+z$ of the uncondensed
particles and the density of the interaction energy,
$(y+v_0x)s$, between the condensate
and the uncondensed particles. Because of the symmetry breaking field, the energy
density of the condensate is still an operator of diagonal and off-diagonal matrix elements
$-\mu s+v_0s^2/2$ and $-\nu\sqrt{s}$, respectively.

For $s>0$ the spectrum of $h(x,y,z,s)$ is purely absolutely
continuous. As a set, it is the interval
$$[\chi(x,y,z,s)-2|\nu|\sqrt{s},\,\chi(x,y,z,s)+2|\nu|\sqrt{s}\,]$$
that can be found by
solving the generalized eigenvalue equation $h\psi=e\psi$ with the
usual exponential ansatz $\psi(m)=\exp(im\alpha)$ and choosing any
real $\alpha$ in the interval $[0,\pi]$. The spectral point corresponding
to $\alpha$ is
\be\label{eal}
e(x,y,z,s,\alpha)=\chi(x,y,z,s)-2\nu\sqrt{s}\cos\alpha\ .
\ee
In particular, the lower edge of the spectrum is
\bea\label{gxyz}
g_{xyz}(\mu,\nu,s)=\frac{v_0}{2}s^2-(\mu-v_0x-y)s-2|\nu|\sqrt{s}\nonumber\\
-\mu x+\frac{v_0}{2}x^2+z
\eea
with two linearly independent solutions $\psi(m)\equiv 1$ and
$\psi(m)=m$. The ground state energy density of $H(\n)$ is
obtained by minimizing the lower edge value with respect to $s$,
\bea\label{exyz}
\lim_{V\to\infty}E_0[H(\n)]/V=e_{xyz}(\mu,\nu)=\min_{s\geq 0}
g_{xyz}(\mu,\nu,s)\nonumber\\
= g_{xyz}(\mu,\nu,s_{xy}(\mu,\nu)),\phantom{aaa}
\eea
where the minimizer $s_{xy}$ does not depend on $z$ \cite{rem4}.
Inspecting the expression
(\ref{gxyz}) we see that whenever $\nu\neq 0$, $s_{xy}>0$ for any value of
$\mu$. Hence, $\partial g_{xyz}(\mu,\nu,s)/\partial s=0$ at $s=s_{xy}$, that is,
$\sqrt{s_{xy}}$ is a positive root of a cubic polynomial.

We have seen in the previous section that in the overall ground state $\n=0$,
therefore $x=y=z=0$ provides the overall ground state energy density:
\bea\label{e0nu}
e_0(\mu,\nu)=\min_{s\geq 0} \{v_0s^2/2-\mu s-2|\nu|\sqrt{s}\,\}\nonumber\\
\equiv\min_{s\geq 0}
g_0(\mu,\nu,s)=g_0(\mu,\nu,s_0(\mu,\nu))\ .
\eea
Equation (\ref{e0nu}) agrees with Eq.~(\ref{8}).
We could have found $e_0$ without knowing the ground state in finite volumes,
by minimizing $e_{xyz}$ over
nonnegative values of $x$, $y$ and $z$.

To obtain Eq. (\ref{7}) we make the following observations:\\
(i) Choosing $\nu\neq 0$,
\begin{widetext}
\bea\label{de0perdnu}
\frac{\partial e_0(\mu,\nu)}{\partial\nu}
=\left[\frac{\partial g_0(\mu,\nu,s)}{\partial\nu}\right]_{s=s_0}
+\frac{\partial s_0(\mu,\nu)}{\partial\nu}
\left[\frac{\partial g_0(\mu,\nu,s)}{\partial s}\right]_{s=s_0}
=\left[\frac{\partial g_0(\mu,\nu,s)}{\partial\nu}\right]_{s=s_0}
=-2\sgn\nu\sqrt{s_0}\ .
\eea
\end{widetext}
(ii) The ground state energy $E_0(\mu,\nu)$ of $H^{\rm FD}_\nu$ is a concave (even) function of $\nu$.
Concavity comes from that of the thermodynamic potential
$-\beta^{-1}\ln\Tr \exp\{-\beta H^{\rm FD}_\nu\}$
by letting the temperature go to zero.
($E_0(\mu,\nu)$ is also an
analytic function of $\nu$ in finite volumes; the singularity of $E_0/V$
at $\nu=0$ develops only as $V$ tends to infinity.)
\\
(iii) The Hellmann-Feynman theorem yields the ground state averages
\be\label{24}
\langle a_0\rangle=\langle a^*_0\rangle=
-\frac{1}{2\sqrt{V}}\frac{\partial E_0(\mu,\nu)}{\partial\nu}\ .
\ee
\\
(iv) According to an inequality known as Griffiths' lemma \cite{Gr} and applied here to the
sequence of concave functions $E_0(\mu,\nu)/V$ converging to $e_0(\mu,\nu)$,
\bea\label{25}
\frac{\partial e_0(\mu,\nu+0)}{\partial\nu}
\leq\liminf_{V\to\infty}\frac{1}{V}\frac{\partial E_0(\mu,\nu)}{\partial\nu}
\phantom{aaaaaaa}
\nonumber\\
\leq\limsup_{V\to\infty}\frac{1}{V}\frac{\partial E_0(\mu,\nu)}{\partial\nu}\leq
\frac{\partial e_0(\mu,\nu-0)}{\partial\nu}
\eea
implying equalities in points where $e_0$ is differentiable. With
(\ref{24}) and (\ref{25}) for $\nu\neq 0$
\be
\alpha_0=-\frac{1}{2}\frac{\partial e_0(\mu,\nu)}{\partial\nu}
\ee
and substituting (\ref{de0perdnu}) into this equation we find Eq. (\ref{7}).

$E_0(\mu,\nu)$ being a concave function also of $\mu$, Eq.~(\ref{ro0}) can be proved from
\be\label{perdmu}
\frac{\partial e_0(\mu,\nu)}{\partial\mu}
=\left[\frac{\partial g_0(\mu,\nu,s)}{\partial\mu}\right]_{s=s_0}=-s_0
\ee
by an identical argument. Because in the Hamiltonian $\mu$ is
coupled to $N$ and not to $N_0$, we can use the result of the former section, that the ground state expectation value
of $N_\k$ vanishes if $\k\neq 0$. Therefore
\be
\rho_0=-\frac{\partial e_0(\mu,\nu)}{\partial\mu}
\ee
which together with (\ref{perdmu}) yields Eq.~(\ref{ro0}).

For $\mu>0$ by solving the cubic equation we find
\be\label{s0}
\sqrt{s_0}=2\sqrt{-q}\,\cos\left(\frac{1}{3}\arctan\frac{\sqrt{-q^3-r^2}}{r}\right)
\ee
with $q=-\mu/3v_0$ and $r=|\nu|/2v_0$.
Expansion of $\sqrt{s_0}$ up to first order in $|\nu|$ yields
\bea\label{1.10}
\alpha_0&=&{\rm sgn\,}\nu \left[\sqrt{\frac{\mu}{v_0}}+\frac{|\nu|}{2\mu}+f_1\nu^2
\right]\nonumber\\
\quad\rho_0&=&\frac{\mu}{v_0}+\frac{|\nu|}{\sqrt{\mu v_0}}+f_2\nu^2\nonumber\\
\quad e_0&=&-\frac{\mu^2}{2v_0}-2|\nu|\sqrt{\frac{\mu}{v_0}}-f_3\nu^2
\eea
where $f_i(\mu,\nu)$ are smooth functions away from zero and bounded at $\nu=0$, and $f_2$ and $f_3$ are positive.

%
\newsec{Gap to one-particle excitations with a nonvanishing momentum}\label{En}

Let us return to Eq. (\ref{exyz}).
The minimum of $g_{xyz}(\mu,0,s)$ is attained at $s=(\mu-\mu_c(x,y))/v_0$ which is
positive if $\mu>\mu_c(x,y)\equiv y+v_0x$.
This is the condition to have a nonvanishing Bose condensate at $\nu=0$
in the ground state of $H(\n)$ with $\n$
chosen so that $x_V(\n)\to x$ and $y_V(\n)\to y$, cf. Eq.~(\ref{xyz}).
If $\nu\neq 0$ and
$\mu\leq \mu_c(x,y)$, the ground state condensate density $s_{xy}$ is still positive but
tends to zero together with $\nu$. Now the full density is $x+s_{xy}$.
Properties
(i)-(iv) listed in the former section remain valid, thus
the order parameter and the condensate density in the ground state of
$H(\n)$ will respectively be
\be\label{alxyz}
\alpha_{xy}=-\frac{1}{2}\frac{\partial\, e_{xyz}(\mu,\nu)}{\partial\nu}
=\sgn\nu\sqrt{s_{xy}}
\ee
and
\be\label{roxyz}
\rho_{xy}=-\frac{\partial\, e_{xyz}(\mu,\nu)}{\partial\mu}-x=s_{xy}.
\ee
We conclude that
the relation $\alpha_{xy}(\mu,\nu)^2=\rho_{xy}(\mu,\nu)$ holds true for
any $\mu$, provided that $\nu\neq 0$.
For $\mu>\mu_c(x,y)$, $s_{xy}$ is given by the right-hand side of Eq.~(\ref{s0})
with $q=-(\mu-\mu_c(x,y))/3v_0$ and $r=|\nu|/2v_0$.

Substituting $s_{xy}$
into Eq.~(\ref{gxyz}) we obtain the ground state energy density
for $H(\n)$. To prove the formula (\ref{10}) this general expression is not
needed.
We set
$\n_\q=1$ and $\n_\k=0$ for $\k\neq \q$.
Then
\be
H(\n)-H(0)=\veps(\q)-\mu+[v_0+v(\q)]N_0/V\ .
\ee
Averaging with the ground states of $H(\n)$ and $H(0)$, respectively,
\bea
\veps(\q)-\mu+[v_0+v(\q)]\langle N_0\rangle_\q/V\phantom{aaaaaaaaaaaaaa}
\nonumber\\
\leq E_\q-E_0
\equiv E_0[H(\n)]-E_0[H(0)]
\nonumber\\
\leq
\veps(\q)-\mu+[v_0+v(\q)]\langle N_0\rangle_0/V\nonumber\\
\eea
by the variational principle. In the present case $x=y=z=0$, so both
$\langle N_0\rangle_\q/V$ and
$\langle N_0\rangle_0/V$ equal $\rho_0+O(1/V)$.
For $\mu>0$ we insert $\rho_0$ from (\ref{1.10}) and obtain Eq.~(\ref{10}).

\newsec{Extension to positive temperatures}\label{pos}

Our analysis of microcanonical ensembles,
especially equations (\ref{alxyz}) and (\ref{roxyz}) make it possible to
extend the result (\ref{Bogogen})
to thermal equilibrium states. Below we outline the argument leading to this
conclusion.

In the microcanonical ensemble $\{x,y,z,s\}$ the condensate density is $s$,
while the order parameter operator
$\lim_{V\to\infty}a_0/\sqrt{V}$ is represented by a matrix $o$ with elements
$o_{m,m+1}=\sqrt{s}$, $-\infty<m<\infty$ and 0 otherwise.
We can compute the
mean value of $o$ in a generalized eigenstate $\psi$ of $h(x,y,z,s)$ by the
formula
\be
\langle o\rangle_\psi=\lim_{M\to\infty}\frac{\sum_{-M\leq m\leq M}
\psi^*(m)(o\psi)(m)}{\sum_{-M\leq m\leq M}|\psi(m)|^2}\ .
\ee
Consider $\psi_{\pm\alpha}(m)=e^{\pm im\alpha}$,
$\psi_{\alpha,c}(m)=\cos m\alpha$ and $\psi_{\alpha,s}(m)=\sin m\alpha$, all
belonging to the spectral point (\ref{eal}). Now
$\langle o\rangle_{\psi_{\pm\alpha}}=e^{\pm i\alpha}\sqrt{s}$,
while
\be\label{oav}
\langle o\rangle_{\psi_{\alpha,c}}=\langle o\rangle_{\psi_{\alpha,s}}=
\cos\alpha\sqrt{s}.
\ee
We want to treat the above
mean values as limits of averages taken with eigenvectors of $H_\nu^{\rm FD}$. By a
general property of one-dimensional second order difference equations, all
eigenvalues of $H(\n)$ are nondegenerate and therefore the eigenvectors are
real. This property is inherited by $H_\nu^{\rm FD}$ if we disregard any possible
accidental coincidence of eigenvalues belonging to different $\n$.
So considering separate eigenvectors the relevant relation is (\ref{oav})
in the sense that for any sequence of eigenvectors $\psi_l$, the volume increasing
with $l$, for all possible limits
\be\label{oav2}
\lim\frac{1}{\sqrt{V}}\frac{\langle\psi_l|a_0|\psi_l\rangle}
{\langle\psi_l|\psi_l\rangle}=\cos\alpha\lim\frac{1}{\sqrt{V}}
\left(\frac{\langle\psi_l|N_0|\psi_l\rangle}{\langle\psi_l|\psi_l\rangle}\right)
^{1/2}
\ee
where $\alpha$ can assume any value. Note also that
computing the average with the help of a density matrix we have to
take a trace and then find the same outcome (\ref{oav}) or (\ref{oav2})
whether we use the complex basis
$\psi_{\pm\alpha}$ or the real one, $\psi_{\alpha,c},\psi_{\alpha,s}$.

Unless $\alpha=0$ or $\pi$, Eq.~(\ref{oav2}) is in conflict with Bogoliubov's
hypothesis, supposing an asymptotic equality
in the Schwarz inequality (\ref{4}).
The fact that Bogoliubov's suggestion proves nevertheless to be correct and we
obtain Eq.~(\ref{Bogogen}) in thermal equilibrium is due to the
strong equivalence of the grand-canonical and a suitably chosen
microcanical ensemble: fixing $\beta=1/k_BT$, $\mu$ and $\nu$ and taking the
thermodynamic limit, the distribution of the intensive variables, $e,x,y,z$
and $s$ among them, becomes a Dirac delta.
Let the energy density be $e_{\beta\mu\nu}$.
Although there is a continuum of the parameter sets $\{x,y,z,s\}$ yielding
$e(x,y,z,s,\alpha)=e_{\beta\mu\nu}$
for different values of $\alpha$, cf. (\ref{eal}), because of the degenerate
limit distribution of these variables only
a unique set will asymptotically dominate. With this set $e_{\beta\mu\nu}$
is attained as the minimum of $e(x,y,z,s,\alpha)$ with respect to $\alpha$ and
$s$ because the density of states of each $H(\n)$
has a maximum at the bottom of the spectrum of $H(\n)$. Minimization
singles out $\alpha=0$ or $\pi$, depending on the sign of $\nu$; see Eq.~(\ref{gxyz}).
In conclusion, $e_{\beta\mu\nu}=e_{xyz}(\mu,\nu)$ and
the thermodynamic limits of $\langle a_0\rangle_{\beta\mu\nu}/\sqrt{V}$ and
$\langle N_0\rangle_{\beta\mu\nu}/V$ are $\alpha_{xy}$ resp. $\rho_{xy}$,
cf. Eqs.~(\ref{exyz}), (\ref{alxyz}) and (\ref{roxyz}),
with $x=x(\beta,\mu,\nu)$, $y=y(\beta,\mu,\nu)$ and $z=z(\beta,\mu,\nu)$.
This proves the validity of Eq.~(\ref{Bogogen}).
Note that $\mu_c(\beta)$, the critical chemical potential of the grand-canonical
ensemble satisfies the relation
\bea
\mu_c(\beta)=\mu_c(x(\beta,\mu_c(\beta),0),y(\beta,\mu_c(\beta),0))\nonumber\\
=y(\beta,\mu_c(\beta),0)+v_0x(\beta,\mu_c(\beta),0).
\eea

Recently Pul\'e and Zagrebnov \cite{PZ}
computed the thermodynamic limit of the pressure of the mean-field Bose gas
in the presence of a
gauge-symmetry breaking field. Their formula reads
\bea
p(\mu,\nu)=\lim_{V\to\infty}\frac{1}{\beta V}\ln\Tr e^{-\beta H^{\rm MF}_\nu}
=\frac{v_0}{2}\rho^2(\mu,\nu)\nonumber\\
+p_{\rm ideal}(\mu-v_0\rho(\mu,\nu))
+\frac{\nu^2}{v_0\rho(\mu,\nu)-\mu}
\eea
where $\rho(\mu,\nu)$ is the unique solution of the equation
\be
\rho=\rho_{\rm ideal}(\mu-v_0\rho)+\frac{\nu^2}{(v_0\rho-\mu)^2}\ .
\ee
Here $p_{\rm ideal}(\mu)$ and $\rho_{\rm ideal}(\mu)$ are respectively
the pressure and density of the ideal Bose gas for $\nu=0$. They are well-defined
if the chemical potential is negative. The authors prove that for $\nu\neq 0$,
$\mu-v_0\rho(\mu,\nu)<0$ indeed. Because the pressure in any finite volume is a
convex function of $\nu$, we can use Griffiths' lemma to obtain
the order parameter from the relation $\alpha_{xy}=\alpha_{x0}
=(1/2)\partial p/\partial\nu$. 
We find $\alpha_{x0}=\nu/(v_0\rho(\mu,\nu)-\mu)$, thus
\bea
\rho_{x0}=\alpha_{x0}^2= 
\rho(\mu,\nu) 
-\rho_{\rm ideal}(\mu-v_0\rho(\mu,\nu)).
\eea
Since $\rho(\mu,\nu)$ is the total density, we can identify
the density of the uncondensed particles as
$$
x=\rho_{\rm ideal}(\mu-v_0\rho(\mu,\nu)).
$$
If $\mu<\mu_c(\beta)=v_0\rho_{\rm ideal}(0)$, $\alpha_{x0}$ goes to zero
with $\nu$:
in this case $\mu-v_0\rho(\mu,\nu)$ converges to a negative value, $\rho(\mu,0)$
is the
unique solution of the equation $\rho=\rho_{\rm ideal}(\mu-v_0\rho)$. This
latter holds true at $\mu=\mu_c(\beta)$ when
$\mu_c(\beta)-v_0\rho(\mu_c(\beta),\nu)$
tends to zero with a vanishing $\nu$, but slower than $\nu$. If
$\mu>\mu_c(\beta)$, the convergence to zero becomes linear and thus the limit
of $\alpha_{x0}$ will be nonzero.

\newsec{Summary}

In this paper we have rigorously established the equivalence of Bose-Einstein
condensation and a spontaneous breakdown of the gauge symmetry in some simplified
models of interacting bosons. In these models one retains only that part of the pair
interaction diagonal in momentum representation.
Equivalence has been
obtained in the form of Eq.~(\ref{Bogogen}),
that we found to hold true for any nonzero value of the symmetry breaking field,
not only in the limit of a vanishing field. As a by-product,
we have disproved the compressibility sum rule in the ground
state of the one-dimensional mean-field Bose gas.
Our method was to study certain
microcanonical ensembles with a fixed density and energy density of uncondensed
particles. In the thermodynamic limit these ensembles are still
described by a family of operators, so-called Jacobi matrices, known from
the one-dimensional tight-binding electron theory.
The limit of infinite volume could be done on the many-body Hamiltonians because
of their commuting with
the occupation number operators of nonzero momentum states; this
made it possible to reduce the problem to that of a single mode, $\k=0$.
The result was first obtained in the
ground state of the microcanonical ensembles and then extended to thermal
equilibrium states by arguing with the strong equivalence of grand- and
microcanonical ensembles.

\newsec*{Acknowledgment}

I am indebted to P\'eter Sz\'epfalusy for many helpful discussions and for
providing me with a copy of Bogoliubov's original report. This work was supported by OTKA Grants T 042914
and T 046129.


\begin{thebibliography}{99}
\bibitem{rem}
Bogoliubov in Ref. \cite{Bog} made the suitable substitutions into his inequality and obtained
the $1/q^2$ singularity necessary to prove the
absence of symmetry breaking in one and two dimensions. To arrive at this conclusion he only should
have summed the resulting inequality over the allowed wave vectors $0<|\q|<q_0$
in finite volumes before letting $V$ tend to infinity (with $q_0$ fixed)
and then $\nu$ to zero.
\bibitem{rem2}
When taking the limit of $E_\q-E_0$ it is understood that $\q$ is replaced by
a sequence $\q_V\to\q$ so that $\q_V$ is allowed for $V$.
\bibitem{rem4}
Because of the convergence in the strong resolvent sense there is a sequence of
eigenvalues of $H(\n)/V$ converging to $e_{xyz}$, and it is easily seen that
$E_0[H(\n)]/V$ cannot tend to a smaller value.
\end{thebibliography}
\end{document}